\documentclass[]{article}
\renewcommand{\arraystretch}{1.3}
\catcode`\@=11
\def\marginnote#1{}

\newcount\hour
\newcount\minute
\newtoks\amorpm
\hour=\time\divide\hour by60
\minute=\time{\multiply\hour by60 \global\advance\minute by-\hour}
\edef\standardtime{{\ifnum\hour<12 \global\amorpm={am}%
        \else\global\amorpm={pm}\advance\hour by-12 \fi
        \ifnum\hour=0 \hour=12 \fi
        \number\hour:\ifnum\minute<10 0\fi\number\minute\the\amorpm}}
\edef\militarytime{\number\hour:\ifnum\minute<10 0\fi\number\minute}

\def\draftlabel#1{{\@bsphack\if@filesw {\let\thepage\relax
      \xdef\@gtempa{\write\@auxout{\string
          \newlabel{#1}{{\@currentlabel}{\thepage}}}}}\@gtempa \if@nobreak
    \ifvmode\nobreak\fi\fi\fi\@esphack} \gdef\@eqnlabel{#1}} \def\@eqnlabel{}
\def\@vacuum{}
\def\draftmarginnote#1{\marginpar{\raggedright\scriptsize\tt#1}}

\def\draft{
  \oddsidemargin -.5truein
  \def\@oddfoot{\footnotesize \sl preliminary draft \hfil
    \rm\thepage\hfil\sl\today\quad\militarytime}
  \let\@evenfoot\@oddfoot \overfullrule 3pt
    \let\label=\draftlabel
    \let\marginnote=\draftmarginnote
  \def\@eqnnum{(\theequation)\rlap{\kern\marginparsep\tt\@eqnlabel}%
    \global\let\@eqnlabel\@vacuum}

  }
\makeatletter
\newdimen\normalarrayskip              % skip between lines
\newdimen\minarrayskip                 % minimal skip between lines
\normalarrayskip\baselineskip
\minarrayskip\jot
\newif\ifold             \oldtrue            \def\new{\oldfalse}
\def\arraymode{\ifold\relax\else\displaystyle\fi} % mode of array entries
\def\eqnumphantom{\phantom{(\theequation)}}     % right phantom in eqnarray
\def\@arrayskip{\ifold\baselineskip\z@\lineskip\z@
     \else
     \baselineskip\minarrayskip\lineskip2\minarrayskip\fi}
\def\@arrayclassz{\ifcase \@lastchclass \@acolampacol \or
\@ampacol \or \or \or \@addamp \or
   \@acolampacol \or \@firstampfalse \@acol \fi
\edef\@preamble{\@preamble
  \ifcase \@chnum
     \hfil$\relax\arraymode\@sharp$\hfil
     \or $\relax\arraymode\@sharp$\hfil
     \or \hfil$\relax\arraymode\@sharp$\fi}}
\def\@array[#1]#2{\setbox\@arstrutbox=\hbox{\vrule
     height\arraystretch \ht\strutbox
     depth\arraystretch \dp\strutbox
     width\z@}\@mkpream{#2}\edef\@preamble{\halign
\noexpand\@halignto
\bgroup \tabskip\z@ \@arstrut \@preamble \tabskip\z@ \cr}%
\let\@startpbox\@@startpbox \let\@endpbox\@@endpbox
  \if #1t\vtop \else \if#1b\vbox \else \vcenter \fi\fi
  \bgroup \let\par\relax
  \let\@sharp##\let\protect\relax
  \@arrayskip\@preamble}
\def\eqnarray{\stepcounter{equation}%
              \let\@currentlabel=\theequation
              \global\@eqnswtrue
              \global\@eqcnt\z@
              \tabskip\@centering
              \let\\=\@eqncr
 \halign to \displaywidth\bgroup
    \eqnumphantom\@eqnsel\hskip\@centering
    $\displaystyle \tabskip\z@ {##}$%
    \global\@eqcnt\@ne \hskip 2\arraycolsep
         $\displaystyle\arraymode{##}$\hfil
    \global\@eqcnt\tw@ \hskip 2\arraycolsep
         $\displaystyle\tabskip\z@{##}$\hfil
         \tabskip\@centering
    &{##}\tabskip\z@\cr}
\begingroup\ifx\undefined\newsymbol \else\def\input#1 {\endgroup}\fi
\newfont{\hr}{msbm10}
\newfont{\ams}{msam10}
\def\2{{1\over 2}}
\def\N2{${\cal N}=2$}
\def\4N{${\cal N}=4$}
\def\1N{${\cal N}=1$}

\def\beq{\begin{equation}}
\def\eeq{\end{equation}}
\def\ba{\beq\new\begin{array}{c}}
\def\ea{\end{array}\eeq}

\voffset= - 1.0in
\hoffset= - 1.0in         % switch off for draft style
\begin{document}
\begin{flushright}
FIAN/TD-03/04\\
ITEP/TH-06/04
\end{flushright}
\vspace{0.5cm}

\begin{center}
\renewcommand{\thefootnote}{${\!}^\star$}
{\Large \bf STRINGS, INTEGRABLE SYSTEMS, GEOMETRY AND
STATISTICAL MODELS
\footnote{%%%%%%
Contribution to the proceedings of the conference
{\em Lie theory and its applications in physics - 5},
June 2003, Varna, Bulgaria } }
\end{center}
\vspace{0.5cm}

\setcounter{footnote}{0}
\begin{center}
{\large A.~Marshakov}\\
%\medskip
\bigskip
{\em Lebedev Physics Institute \&
ITEP, Moscow, Russia}\\
\medskip
{\sf e-mail:\ mars@lpi.ru, mars@itep.ru}\\
\bigskip\bigskip\medskip
\end{center}

\begin{quotation}
\noindent
The role of integrable systems in string theory is discussed.
We remind old examples of the correspondence between
stringy partition functions or effective actions
and integrable equations, based on effective application of the
matrix model technique. Then we turn to a new example, coming from the
Nekrasov deformation of the Seiberg-Witten prepotential. In the last case
the deformed theory is described by a different statistical model,
which becomes equivalent to a partition
function of a topological string. The full partition function of string
theory arises therefore always as a certain "quantization" of its
quasiclassical geometry.
\end{quotation}

\section{Introduction}

There is lack of algorithmic direct computational
methods in non-perturbative physics. One of the
working ideas is to replace the divergent series of perturbation theory by
differential equations in some parameter space (say, in the space of
couplings) and to study their solutions, which can be in many cases treated
non-perturbatively.
Another way is based on some discrete (e.g. lattice) regularization of
continuous theory, which in practice turn it into some statistical model,
where one may hope to reach more progress.
Amazingly enough these two ways are closely related, at least in the class
of models, where the exact non-perturbative results have been reached during
last fifteen years, (see e.g.~\cite{np}, where these issues were discussed
from different angles of view).

Our experience shows, that general scheme of formulation of the
non-perturbative results is always the same: the generating
function $F=F(t)$ should be identified with a solution to a hierarchy
of nonlinear differential equations and, when the system is known,
is given by a logarithm of tau-function of this
hierarchy. The string theory genus expansion
\begin{equation}
\label{expan}
F = \sum_{g=0}^\infty \hbar^{2g-2}F_{g}
\end{equation}
is then equivalent to the so-called {\em quasiclassical}
expansion of $F$ and therefore we denote the string coupling as
a Planck constant $\hbar$.

The leading term of this expansion $F_0=F_0(t,S)$,
(depending usually on the set of extra parameters or moduli $S$)
is already well-known for all examples and can be defined
in pure {\em geometric} terms. The geometric data contain a
complex manifold (in practically all studied cases
a curve) $\Sigma$ endowed
with pair of functions $(x,y)$, or with the generating
meromorphic form $dS=ydx$, then $F_0$ can be identified with the
Krichever quasiclassical tau-function \cite{KriW}, sometimes also called
a "prepotential".

The whole sum in (\ref{expan}) can be thought of as a "quantization" of this
geometry. Except for few simple examples (the most well-known is Kontsevich
model and its generalizations \cite{K,GKM,other}) yet there is
no strict formulation
of the full function $F$, but it is widely beleived, that
\begin{equation}
\label{ZF}
Z = \exp\left(- F \right)
\end{equation}
satisfies some (infinite system of)
relatively simple linear differential equations, like Virasoro or
$W$-constraints \cite{vir} together with nonlinear equations of an
integrable hierarchy.

The pair of functions $(x,y)$ entering the geometric data for $F_0$ may be
thought of as "symplectic pair", and different solutions (their
Baker-Akhiezer functions) are related by
a sort of Fourier transform \cite{pqdual}
\begin{equation}
\label{pqdual}
\Psi_i (y) \propto \int d\mu(x) \Psi_i (x) e^{\int ydx}
\end{equation}
In particular, this is the origin of existence of
integral representations of the solutions to
Kontsevich-like models since by (\ref{pqdual})
they are "dual" to trivial tau-functions. The corresponding partition
functions
\begin{equation}
\label{detfo}
Z\left(t_k = -{1\over k}\sum x_j^{-k}\right) \propto
{\det \Psi_i(x_j)\over\Delta(x)}
\end{equation}
are given by matrix version of (\ref{pqdual}), since for trivial
$\Psi$-functions in the r.h.s. formula (\ref{pqdual})
gives rise to the matrix integral representation of $Z$.
However, these exact full answers are known only for the cases when
quasiclassical geometry is extremely simple, or, basically, when
$\Sigma$ is
sphere. Already for the $(p,q)$ critical points of two-dimensional gravity
(with $p,q\neq 1$; $(p,1)$ points correspond to generalized Kontsevich
models \cite{GKM})
generic curves are of genus ${1\over 2}(p-1)(q-1)$ \cite{pq1}
and nothing strictly is known about explicit form of the full $F$.
Recently these problems were readdressed to in \cite{SeVa} and from
a different angle of view in \cite{amm}.

Hence, the quasiclassical part of the story looks like being totally based
on geometry. However, its "quantization" leads to relatively simple
statistical models, like matrix integrals. The present hope is that this is
a general phenomenon, and as a new example of this phenomenon
we will consider the
Nekrasov deformation \cite{Nekrasov} of the Seiberg-Witten theory, which
leads to "Lie-algebraic construction" of the full partition function
based on summing over the Young diagrams.

\section{Matrix models and integrable systems}

Matrix models are the most simple example of the
"gauge-string correspondence" (\ref{ZF})
where $\log Z=F$ obeys (\ref{expan}). A typical matrix model
partition function is given by an integral
\begin{equation}
\label{mamo}
Z = \int d\Phi e^{-{1\over\hbar}{\rm Tr} V(\Phi)} =
{1\over N!}\int \prod_{i=1}^N \left(dx_i
e^{-{1\over\hbar}V(x_i)}\right)
\Delta^2(x)
\end{equation}
where $\Phi$ is $N\times N$ matrix (in general situation the integral
is taken over ensemble of matrices), for example Hemitean, the measure
$d\Phi\propto\prod d\Phi_{ij}$ and potential $V(\Phi)= \sum t_k\Phi^k$
in this case is an arbitrary polynomial. The simple matrix integral like
(\ref{mamo}) can be rewritten as an integral over eigenvalues
$x_i$ of the matrix $\Phi$. The full partition function
(\ref{mamo}) can be considered as a "quantization" of the
geometry, corresponding to its limit $N\to\infty$, $\hbar\to 0$, with
$N\hbar=t_0={\rm fixed}$.

In this quasiclassical limit $\hbar\to 0$ the first term $F_0$ of the
expansion (\ref{expan}) for partition function (\ref{mamo})
can be determined from a variational problem
\begin{equation}
\label{variF}
F_0 = \left[\int V\rho -
\int\int \rho_1\log\left|x_1-x_2\right|\rho_2
+ \sum_k \Pi_k \left(\int_{D_k}\rho - S_k\right)
\right]_{{\delta F_0\over\delta\rho}=0}
\end{equation}
where $S_k$ are the new quasiclassical parameters -- the
fractions of eigenvalues on various supports $D_k$, and
$\Pi_k$ are corresponding Lagrange multiplyers.
This function can be described in pure geometric terms as a {\em
prepotential}, i.e. the variational
problem (\ref{variF}) is solved by $F_0={\cal F}(t,S)$, where
\begin{equation}
\label{dF}
{\partial {\cal F}\over\partial S_i} = \Pi_i = \oint_{B_i} dS,
\ \ \ \ \
{\partial {\cal F}\over\partial t_k} = {\rm res}\ (x^kdS)
\end{equation}
with the matrix model geometry appearing in the following way.

Loop equations \cite{Migdal}
in the main order define the complex curve of the matrix
model $\Sigma$, which is
of genus $g_1=({\rm deg}V-2)$ for the 1-matrix model \cite{David}
(where always hyperelliptic),
and of (now always non-hyperelliptic) of
genus $g_2=({\rm deg}V_1-1)({\rm deg}V_2-1)-1$
for the 2-matrix model with potential $V=-xy+V_1(x)+V_2(y)$
\cite{KM}. The density of eigenvalues $\rho$
endows $\Sigma$ with
a meromorphic 1-form $dS$, so that the fractions of eigenvalues are given
by its periods $S_i = \oint_{A_i} dS$ on $\Sigma$.

For the symmetric 2-matrix model the matrix integral
(\ref{mamo}) and the variational problem (\ref{variF}) become now
two-dimensional, so that the inverse to the kernel in (\ref{variF}) is
simply a two-dimensional Laplace operator.
In this way the problem for 2-matrix model can be reformulated as a
potential problem
for two-dimensional domains in complex plane, some aspects of which
are sometimes referred to as "Laplacian growth" of domains w.r.t.
$t_0=N\hbar$
with fixed coefficients $t$ of the potential $V$,
(see, e.g. \cite{ragro} and references therein).
In the simplest case this problem is solved
in terms of the dispersionless Toda lattice tau-function
(the corresponding geometric data and prepotential
are sometimes referred to as tau-functions of curves \cite{taucur}),
but for many
domains the equations are much less trivial, their explicit form
in this case is derived in \cite{KMZ}.

The simplicity of dispersionless solution comes from the fact that curve
$\Sigma$ for this case is just sphere, which allows to
express differentials $d\Omega_k={\partial
dS\over \partial t_k}$ in terms of rational functions of
the uniformizing global co-ordinate $w=e^{\Omega_0}$
as $\Omega_k = P_k(w)$. This
gives rise to dispersionless Hirota equations on the (second derivatives of)
$F_0$. An alternative way
is to write dispersionless $W$-constraints of the form
$\sum{\rm res} \left(x^ny^mdx\right) = 0$.

In all known examples the dispersionless $W$-constraints are reminisents of
the $W$-constraints \cite{vir} imposed onto the full partition function
(\ref{ZF}).
It is possible then to define the full partition function (\ref{ZF}) as a
correlator of some free fields on $\Sigma$, bosons or fermions
$\tilde\psi_k=e^{i\phi_k}$, $\psi_k=e^{-i\phi_k}$
\cite{conf}, (see also \cite{Kostov}).
The matrix integral (\ref{mamo}) is an example of such correlator
$Z=\left< \exp\left(\oint V(x)\partial\phi(x)\right)
G\right>$ with a very simple "element of Grassmanian"
$G=\exp\left(\int e^{\sqrt{2}\phi}\right)$ commuting with the $W$-generators,
but for less trivial cases, like in (\ref{detfo}),
the explicit representation of $G$ in terms
of bosons or fermions is not yet known.

Note, that the matrix integrals (\ref{mamo}) from string theory point of
view correspond to the gauge or open string theory, in contrast to the
Kontsevich like integrals, which are rather simplest examples of the closed
string field theory \cite{sft}. For example, for the gaussian potential
$V(\Phi)\propto\Phi^2$ both descriptions are known \cite{todagkm} which results
in a nontrivial identity for two integrals over matrices of different sizes.

\section{Seiberg-Witten geometry, Young diagrams and topological
strings\label{ss:young}}

Another example, now of more realistic theory, where geometry
determines $F_0$ is the famous Seiberg-Witten theory \cite{SW}, or
the exact solution for the low-energy prepotential
of the $N=2$ SUSY four-dimensional Yang-Mills vector multiplet
with the gauge field $A$, complex scalar $\Phi$ and adjoint fermions.
The geometric data
again contain complex curve $\Sigma $ of finite genus,
endowed with a meromorphic
differential $dS$ with special properties or a finite-gap integrable system.
The Seiberg-Witten prepotential
is introduced by analog of (\ref{dF})
\begin{equation}
{\partial {\cal F}^{SW}\over\partial a_i} =
%a^D_i=
\oint_{B_i}dS
\end{equation}
which is equivalent to ${\partial^2{\cal F}\over\partial a_i\partial a_j} = T_{ij}$,
where $T_{ij}$ is period matrix of $\Sigma$, defined together with
$dS$ by equations
\begin{equation}
\label{suncu}
w + {\Lambda^{2N}\over w} = P_{N}(\lambda),
\ \ \ \
dS = ydx\equiv\lambda{dw\over w},
\ \ \ \
S_i = a_i\equiv\oint_{A_i}dS
\end{equation}
i.e. $F_0={\cal F}^{SW}(a,\Lambda)$ is again a
quasiclassical tau-function \cite{GKMMM}. The role of quasiclassical
parameters $S$ is now played by the condensates (BPS masses) $a$
(in addition to $t_0=\tau_0\propto\log\Lambda$)
and generating parameters
$t$ can be introduced in a
standard way \cite{RG} along the general lines of \cite{KriW}.

From the example with matrix models we have learnt that the "quantization"
of the matrix model geometry is given in terms of a relatively simple
statistical model. One may ask, what is the analog of matrix model for the
Seiberg-Witten geometry, and what is the physical sense of corresponding
"quantization"? An answer to this question was proposed by
Nekrasov \cite{Nekrasov}, which leads to different from integrals over
matrices statistical model.

One should consider the {\em deformation} of the Seiberg-Witten theory by
putting it into nontrivial supergravity background with twisted boundary
conditions on the fields. A natural way is to use the Scherk-Schwarz mechanism
in six-dimensional theory on ${\bf R}^4 \times {\bf T}^2$ with metric
$ds^2 = r^2 dz d{\bar z}+(dx^{\mu}+2{\rm Re} ({\Omega}^{\mu}_{\nu}x^{\nu} dz))^2$
and an extra Wilson loop ${\bf A}(\Omega)$,
where holonomy matrix $\| \Omega_\mu^\nu\| =
{\rm diag}(i\epsilon_1,i\epsilon_2)\otimes\sigma_2$.

Another part of deformation is topological twisting of the SUSY algebra,
and the {\em holomorphic} perturbation of the effective action,
which comes from the perturbation of the ultraviolet action by the
operators ${{\rm Tr}} {\Phi}^J$ (and their products)
provided the conjugated "prepotential" is kept classical
${\bar\tau}_0 {{\rm Tr}} {\bar\Phi}^2$.
This leads to the deformation of the Seiberg-Witten prepotential
by holomorphic operators
${\cal O}_{\vec n}(a) = \left< \prod_{J=1}^{\infty}{1\over
n_{J}!}  \left( {1\over J}{{\rm Tr}} {\Phi}^{J} \right)^{n_J}
\right>_{\langle\Phi\rangle = a}$
\begin{equation}
{{\cal F}} ( a, {\tau}_{\vec n} ) =
{{\cal F}}^{SW} (a ; {\tau}_0 ) + \sum_{\vec n} {\tau}_{\vec n}
{\cal O}_{\vec n} (a) +
 \sum_{\vec n, \vec m } {\tau}_{\vec n}
{\tau}_{\vec m} {\cal O}_{\vec n\vec m} (a) + \ldots
\end{equation}
where ${\cal O}_{\vec n\vec m} (a)$ are the "contact terms".

It was shown in \cite{Nekrasov,LMN}, that computation of so deformed
deformed partition function leads to statistical model of
"random growth" of the Young diagrams. The standard argument with
$Q_{BRST}$-exact terms allows
computation of the chiral observables at ${\bar\tau}_0 \to \infty$
so that the term ${\bar\tau}_0 \Vert F_A^{+} \Vert^2$ leads to
the self-duality constraint $F_A^{+} = 0$ and the partition function
\begin{equation}
Z({\tau}_{\vec n}; a, {\Omega}) =
\int_{{\phi}({\infty}) = a} D{\Phi}DA \ldots
e^{-S({\Omega})}
\end{equation}
localizes onto instanton solutions, described by the ADHM "matrix model".
As a result, the
localization on the fixed points in the moduli space
${\cal M} = \{F_A^+=0\}/{\cal G}$ gives rise to a
statistical model with summing taken over partitions $\vec{\bf k}$
\begin{equation}
Z = \int_{\cal M} d\mu
\exp\left(-\sum_{\vec n}\tau_{\vec n}{\cal O}_{\vec n}\right) =
 \sum_{\vec{\bf k}}\mu_{\vec{\bf k}}
\exp\left(-\sum_{J}t_J{\cal O}_{J}[{\vec{\bf k}}]\right)
\end{equation}
with the chiral operators replaced by quantities
$$
{\cal O}_{\vec n}(\Phi) \rightarrow {\cal O}_J [{\vec{\bf k}}] =
 {1\over J} \sum_{l, i}
\left[ \left( (a_l + {\hbar} (k_{li} + 1- i) \right)^J - \left(
a_l + {\hbar} (k_{li} - i)\right)^J \right]
$$
In the $U(1)$ case \cite{LMN}
one gets the sum over the Young diagrams, where, due to the "hook formulas"
$$
{\mu}_{\bf k} = (-1)^{k}\left[
\prod_{i<j}^n{\left({\hbar} \left(k_i-k_j+j-i\right)\right)}
\prod_{i=1}^n{1\over {\hbar}^{k_i + n -i}
(k_i+n-i)!} \right]^2 =
(-1)^k \left[ {{\rm dim}R_{\bf k}
\over {\hbar}^k \ k!} \right]^2
$$
and summing using Burnside's theorem we get
(with $t_K=0$ for $K>1$)
\begin{equation}
Z = e^{-t_1
{{a^2}\over 2{\hbar}^2}} \sum_{\bf k} {\mu}_{\bf k} e^{t_1 \vert {\bf k}
\vert}
 = \exp\left[ - {1\over {\hbar}^2} \left( t_1 {a^2
\over 2} + e^{t_1} \right)\right]
= \exp \left(-{F\over {\hbar}^2}\right)
\end{equation}
where $F=F_0$ (in this case exactly!) is the partition functon of the
topological string on ${\bf CP}^1$!

To sum effectively over partitions one may apply fermionic
formalism of integrable systems, see for example \cite{Kharchev}.
For the generating function of all chiral
operators in $U(1)$ theory we get \cite{LMN}
\begin{equation}
Z = {{\left< {a\over\hbar} \left| e^{J_{1} \over
{\hbar}} {\exp} \left[  \sum_{p = 1}^{\infty} {\hat t}_{p} W_{p+1}
\right] e^{-{J_{-1}\over{\hbar}}} \right| {a\over\hbar} \right>}}
\end{equation}
where $W$-background is introduced by the formulas
$$
W_{p+1}
= {1\over \hbar} \oint \ : \ {\widetilde \psi} \left( {\hbar} D
\right)^{p} {\psi}, \ \ \ \  D = z{\partial}_z
$$
$$
\sum_{p=1}^{\infty} {\hat t}_{p} \ x^{p} =
\sum_{p=1}^{\infty} {t_{p}\over{(p+1)!}} {{(x+{\hbar \over
2})^{p+1} - (x - {\hbar \over 2})^{p+1}}\over \hbar}
$$
As in matrix model case, this is the tau-function
of the Toda lattice hierarchy, which can be checked \cite{LMN} by direct
comparison with \cite{OP}, after applying a Bogolyubov transformation.

For the $U(N)$ theory it was shown in \cite{NO} that growth of random
partitions or
Young diagrams (which replaces here the Laplacian growth of the domains
of eigenvalues) leads to appearence of the Seiberg-Witten curves, which
generalize the Vershik-Kerov asymptotic curves for the Young diagrams.

\section{Discussion}

In this notes we have concentrated on the geometric properties of the
quasiclassical exact string theory solutions and their quantization which
leads to representation of stringy partition functions in terms
of simple statistical models.

The quasiclassical part is relatively well-known, and it
always has a geometric formulation. The situation
with its full or "quantum" analog is rather more poor.
Already the first correction to $F_0$ in (\ref{expan}) or $F_1$ is
much less studied (see, however the computations of \cite{Ak,Ey,Ch} and
discussion of this issue in \cite{Kostov,D,EKoKo}), though there is a
common beleif that it is determined by determinant of the
Dirac $\bar\partial$-operator on $\Sigma$,
i.e. $F_1=-{1\over 24}\log\det \bar\partial_\Sigma$.
As for the full function $F$ it is not known
(and even hardly beleived to be possible) how to define it in
terms of geometry on $\Sigma$.

It is necessary to point out that already the sense of
$S$-variables, naturally appearing in (\ref{variF}) due to condensates
of eigenvalues and corresponding to possible nontrivial form of $\Sigma$,
is unclear for
the full "quantum" theory (\ref{mamo}). In particular cases, when the full
solution is known, the dependence on these parameters is almost trivial
\cite{LGGKM}. There is a clear understanding that dispersionless geometry of
sphere is "quantized" into solutions of the ordinary KP or Toda hierarchies,
but it is not at all clear what is the hyputhetical "dispersional"
analog of generic
quasiclassical integrable systems of \cite{KriW}.

The example of sect.~\ref{ss:young} shows also that the nature of
quasiclassical
expansion can be rather nontrivial for interesting physical theories, and
for the supersymmetric gauge theory it is not naive and
corresponds to some dual topological
string. This is a new interesting example of gauge/string duality, which
nevertheless fits in the general scheme we discussed above.
The "quantization" of the Seiberg-Witten geometry leads in this way to
appearence of different from known before in this context
statistical model (other examples of statistical models arising in a
similar way were recently considered in
\cite{VafOR,VafNekIqbal}). However, all them concern only the
topological string
models. One may hope, nevertheless, that analogous phenomena occur even for
the "realistic" four-dimensional gauge theories, and the first sign of
this is the appearence of the algebro-geometric
solutions to spin chains \cite{MZ} in study of renormalization of operators
in $N=4$ SUSY Yang-Mills. These solutions correspond to the classical
approximation in dual string theory, and one may try to understand the
"quantization" of corresponding geometry, discussed above, in terms of
usual physical quantization of corresponding string theory.

\section*{Acknowledgments}

I am indebted to I.Krichever, A.Losev, A.Mironov,
A.Morozov and N.Nekrasov for very important discussions, and
I am grateful to Vlado Dobrev and other organizers of the workshop
{\em Lie theory and its applications in physics - 5} for the
warm hospitality in Varna. The work was partially supported by RFBR grant
02-02-16496, the grant for support of scientific schools 1578.2003.2,
and the Russian Science Support Foundation.

%%%%%%%%%%%%%%%%%%%%%%%%%%%%%%%%%%%%%%%%%%%%%%%%%%%%%%%%%%%%%
% Doing Appendix(ices)                                      %
%%%%%%%%%%%%%%%%%%%%%%%%%%%%%%%%%%%%%%%%%%%%%%%%%%%%%%%%%%%%%

%\appendix

%\section{HEADING FOR APPENDIX A}

%\renewcommand{\theequation}{A.\arabic{equation}}

%TYPE TEXT FOR APPENDIX A HERE.

%\section{HEADING FOR APPENDIX B}

%\renewcommand{\theequation}{B.\arabic{equation}}

%TYPE TEXT FOR APPENDIX B HERE.

%%%%%%%%%%%%%%%%%%%%%%%%%%%%%%%%%%%%%%%%%%%%%%%%%%%%%%%%%%%%%
% Doing references:                                         %
%%%%%%%%%%%%%%%%%%%%%%%%%%%%%%%%%%%%%%%%%%%%%%%%%%%%%%%%%%%%%


\begin{thebibliography}{0}

%%%%%%%%%%%%%%%%%%%%%%%%%%%%%%%%%%%%%%%%%%%%%%%%%%%%%%%%%%%%%
%                                                           %
% Command to used is:-                                      %
%                                                           %
%  \bibitem{REFERENCE_LABEL} AUTHORS NAMES,                 %
%  {\it JOURNAL'S NAMES}{\bf VOLUME NUMBER}, PAGE (YEAR).   %
%                                                           %
%  See example below.                                       %
%                                                           %
%%%%%%%%%%%%%%%%%%%%%%%%%%%%%%%%%%%%%%%%%%%%%%%%%%%%%%%%%%%%%

\bibitem{np}
V.~G.~Knizhnik,
 %``Multiloop Amplitudes In The Theory Of Quantum Strings And Complex
%Geometry,''
Sov.\ Phys.\ Usp.\  {\bf 32}, 945 (1989);\\
A.~Morozov,
%``String Theory: What Is It?,''
Sov.\ Phys.\ Usp.\  {\bf 35}, 671 (1992);\\
R.~Dijkgraaf,
%``Intersection theory, integrable hierarchies and topological field theory,''
arXiv:hep-th/9201003;\\
M.~Bershadsky, S.~Cecotti, H.~Ooguri and C.~Vafa,
 %``Kodaira-Spencer theory of gravity and exact results for quantum string
%amplitudes,''
Commun.\ Math.\ Phys.\  {\bf 165}, 311 (1994)
[arXiv:hep-th/9309140];\\
J.~Polchinski,
%``What is string theory?,''
arXiv:hep-th/9411028;\\
A.~Marshakov,
in {\sl Integrable models and strings}, Lectures
at 3rd Baltic Student Seminar, Helsinki, 1993,
%``String theory and classical integrable systems,''
[arXiv:hep-th/9404126];
{\sl Seiberg-Witten Theory and Integrable Systems},
World Scientific, 1999;
%``String theory or field theory?,''
Phys.\ Usp.\  {\bf 45}, 915 (2002)
%[Usp.\ Fiz.\ Nauk {\bf 45}, 977 (2002)]
[arXiv:hep-th/0212114].
%
\bibitem{KriW}I.~Krichever,
Commun. Pure. Appl. Math. {\bf 47} (1994) 437, hep-th/9205110.
%
\bibitem{K}
%М.Концевич, Функ.Ан. и Прилож. {\bf 25} (1991) 50;\\
M.Kontsevich, Func.~Anal.\& Apps. {\bf 25} (1991) 50;
Comm. Math.Phys. {\bf 147} (1992) 1.
%
\bibitem{GKM}
S.~Kharchev, A.~Marshakov, A.~Mironov, A.~Morozov and A.~Zabrodin,
%``Unification of all string models with C < 1,''
Phys.\ Lett.\ B {\bf 275}, 311 (1992)
[arXiv:hep-th/9111037];
%``Towards unified theory of 2-d gravity,''
Nucl.\ Phys.\ B {\bf 380}, 181 (1992)
[arXiv:hep-th/9201013].
%
\bibitem{other}
M.~Adler and P.~van Moerbeke,
%``A Matrix Integral Solution To Two-Dimensional W(P) Gravity,''
Commun.\ Math.\ Phys.\  {\bf 147}, 25 (1992);\\
C.~Itzykson and J.~B.~Zuber,
%``Combinatorics Of The Modular Group. 2. The Kontsevich Integrals,''
Int.\ J.\ Mod.\ Phys.\ A {\bf 7}, 5661 (1992)
[arXiv:hep-th/9201001].
%
\bibitem{vir}
M.Fukuma, H.Kawai and R.Nakayama,
Int.J.Mod.Phys. {\bf A6} (1991) 1385-1406;\\
%
R.Dijkgraaf, H.Verlinde and E.Verlinde,
Nucl.Phys. {\bf B348} (1991) 435;\\
%
A.Gerasimov, A.Marshakov, A.Mironov, A.Morozov and A.Orlov,
Nucl.Phys. {\bf B357} (1991) 565-618;\\
%
E.~Witten,
%``On the Kontsevich model and other models of two-dimensional gravity,''
IASSNS-HEP-91-24;\\
%
A.~Marshakov, A.~Mironov and A.~Morozov,
%``On equivalence of topological and quantum 2-d gravity,''
Phys.\ Lett.\ B {\bf 274}, 280 (1992)
[arXiv:hep-th/9201011];\\
%
T.~Eguchi, K.~Hori and C.~S.~Xiong,
%``Gravitational quantum cohomology,''
Int.\ J.\ Mod.\ Phys.\ A {\bf 12}, 1743 (1997)
[arXiv:hep-th/9605225].
%
\bibitem{pqdual}
S.~Kharchev and A.~Marshakov,
 %``Topological versus nontopological theories and p - q duality in c <= 1 2-d
%gravity models,''
in
{\sl String Theory, Quantum Gravity and the Unification of the Fundamental
Interactions}, Proceedings of
International Workshop on String Theory, Quantum Gravity and the
Unification of Fundamental
Interactions, Rome, Italy, 1993,  World Scientific;
arXiv:hep-th/9210072;
%``On p - q duality and explicit solutions in c <= 1 2-d gravity models,''
Int.\ J.\ Mod.\ Phys.\ A {\bf 10}, 1219 (1995)
[arXiv:hep-th/9303100].
%
\bibitem{pq1}
I.Krichever, {\sl On Heisenberg relations for the ordinary linear
differential operators}, preprint ETH (1990);\\
%С.Новиков, {\sl Квантование конечнозонных потенциалов и нелинейная
%квазиклассика, возникающие в непертурбативной теории струн},
%Функ.Ан. и Прилож. {\bf 24} (1990) 43-53.
S.Novikov, Func.~Anal.\& Apps. {\bf 24} (1990) 43-53;\\
R.Schimmrigk,
%{\sl Geometric formulation of nonperturbative 2D quantum gravity},
Phys.Rev.Lett. {\bf 65} (1990) 2483-2486;\\
G.Moore,
%{\sl Geometry of string equations},
Comm.Math.Phys. {\bf 133}(1990) 261-304;\\
A.~Marshakov,
%``Exact solutions to quantum field theories and integrable equations,''
Mod.\ Phys.\ Lett.\ A {\bf 11}, 1169 (1996)
[arXiv:hep-th/9602005].
%
\bibitem{SeVa}
M.~Aganagic, R.~Dijkgraaf, A.~Klemm, M.~Marino and C.~Vafa,
%``Topological strings and integrable hierarchies,''
arXiv:hep-th/0312085;\\
N.~Seiberg and D.~Shih,
%``Branes, rings and matrix models in minimal (super)string theory,''
arXiv:hep-th/0312170.
%
\bibitem{amm}
A.~Alexandrov, A.~Mironov and A.~Morozov,
 %``Partition functions of matrix models as the first special functions of
%string theory. I: Finite size Hermitean 1-matrix model,''
arXiv:hep-th/0310113.
%
\bibitem{Nekrasov}
N.~Nekrasov,
%``Seiberg-Witten prepotential from instanton counting,''
arXiv:hep-th/0206161.
%
\bibitem{Migdal}
A.~A.~Migdal,
%``Loop Equations And 1/N Expansion,''
Phys.\ Rept.\  {\bf 102}, 199 (1983).
%
\bibitem{David}
F.~David,
%"Non-Perturbative Effects in Matrix Models and Vacua of Two Dimensional Gravity",
Phys.Lett. B302 (1993) 403-410, hep-th/9212106;\\
G.~Bonnet, F.~David, B.~Eynard,
%``Breakdown of universality in multi-cut matrix models,''
J.Phys. {\bf A33} (2000) 6739-6768, cond-mat/0003324.
%
\bibitem{KM}
V.~Kazakov and A.~Marshakov,
%``Complex curve of the two matrix model and its tau-function,''
J.\ Phys.\ A {\bf 36}, 3107 (2003)
arXiv:hep-th/0211236.
%
\bibitem{ragro}
M.~Mineev-Weinstein, P.~Wiegmann and A.~Zabrodin,
Phys.Rev.Lett. 84 (2000) 5106-5109
arXiv:nlin.SI/0001007;\\
I.~Krichever, M.~Mineev-Weinstein, P.~Wiegmann and A.~Zabrodin,
nlin.SI/0311005.
%
\bibitem{taucur}
P.~Wiegmann and A.~Zabrodin,
%``Conformal maps and dispersionless integrable hierarchies,''
Commun.\ Math.\ Phys.\  {\bf 213}, 523 (2000)
[arXiv:hep-th/9909147];\\
I.~K.~Kostov, I.~Krichever, M.~Mineev-Weinstein, P.~B.~Wiegmann and A.~Zabrodin,
%``$\tau$-function for analytic curves,''
arXiv:hep-th/0005259;\\
%
A.~Marshakov, P.~Wiegmann and A.~Zabrodin,
%``Integrable structure of the Dirichlet boundary problem in two  dimensions,''
Commun.\ Math.\ Phys.\  {\bf 227}, 131 (2002)
[arXiv:hep-th/0109048].
%
\bibitem{KMZ}
I.~Krichever, A.~Marshakov and A.~Zabrodin,
 %``Integrable structure of the Dirichlet boundary problem in multiply-connected
%domains,''
arXiv:hep-th/0309010;\\
A.~Marshakov and A.~Zabrodin,
%``On the Dirichlet boundary problem and Hirota equations,''
arXiv:hep-th/0305259.
%
\bibitem{conf}
A.~Marshakov, A.~Mironov and A.~Morozov,
%``Generalized matrix models as conformal field theories: Discrete case,''
Phys.\ Lett.\ B {\bf 265}, 99 (1991);\\
S.~Kharchev, A.~Marshakov, A.~Mironov, A.~Morozov and S.~Pakuliak,
 %``Conformal matrix models as an alternative to conventional multimatrix
%models,''
Nucl.\ Phys.\ B {\bf 404}, 717 (1993),
arXiv:hep-th/9208044.
%
\bibitem{Kostov}
I.~K.~Kostov,
%``Conformal field theory techniques in random matrix models,''
arXiv:hep-th/9907060.
%
\bibitem{sft}
A.~Marshakov,
%``On string field theory for C <= 1,''
in Proc. of 16th Johns Hopkins Workshop on
Current Problems in Particle Theory: Pathways to Fundamental Theories,
arXiv:hep-th/9208022.
%
\bibitem{todagkm}
S.~Kharchev, A.~Marshakov, A.~Mironov and A.~Morozov,
 %``Generalized Kontsevich model versus Toda hierarchy and discrete matrix
%models,''
Nucl.\ Phys.\ B {\bf 397}, 339 (1993)
[arXiv:hep-th/9203043].
%
\bibitem{SW}
N. Seiberg and E. Witten,
Nucl.Phys., {\bf B426} (1994) 19-52;
{\bf B431} (1994) 484-550.
%
\bibitem{GKMMM}
A. Gorsky, I. Krichever, A. Marshakov, A. Mironov and A. Morozov,
Phys.Lett. {\bf B355} (1995) 466-477, hep-th/9505035.
%
\bibitem{RG}
A. Gorsky, A. Marshakov, A. Mironov and A. Morozov,
%"RG flows from Whitham hierarchy" hepth/9802004
Nucl.Phys., {\bf B527} (1998) 690-716, hep-th/9802004.
%
\bibitem{LMN}
A.~S.~Losev, A.~Marshakov and N.~Nekrasov,
%``Small instantons, little strings and free fermions,''
arXiv:hep-th/0302191.
%
\bibitem{Kharchev}
S.~Kharchev, hep-th/9810091.
%
\bibitem{OP}
A.~Okounkov, R.~Pandharipande, math.AG/0207233,
math.AG/0204305.
%
\bibitem{NO}
N.~Nekrasov and A.~Okounkov,
%``Seiberg-Witten theory and random partitions,''
arXiv:hep-th/0306238.
\bibitem{Ak}
G.~Akemann,
 %``Higher genus correlators for the Hermitian matrix model with multiple
%cuts,''
Nucl.\ Phys.\ B {\bf 482}, 403 (1996)
[arXiv:hep-th/9606004]
%
\bibitem{Ey}
B.~Eynard,
%``Large N expansion of the 2-matrix model,''
JHEP {\bf 0301}, 051 (2003)
[arXiv:hep-th/0210047].
%
\bibitem{Ch}
L.~Chekhov,
%``Genus one correlation to multi-cut matrix model solutions,''
arXiv:hep-th/0401089.
%
\bibitem{D}
A.~Klemm, M.~Marino and S.~Theisen,
 %``Gravitational corrections in supersymmetric gauge theory and matrix
%models,''
JHEP {\bf 0303}, 051 (2003)
[arXiv:hep-th/0211216];\\
R.~Dijkgraaf, A.~Sinkovics and M.~Temurhan,
%``Matrix models and gravitational corrections,''
arXiv:hep-th/0211241.
%
\bibitem{EKoKo}
B.~Eynard, A.~Kokotov and D.~Korotkin,
%``$1/N^2$ correction to free energy in hermitian two-matrix model,''
arXiv:hep-th/0401166.
%
\bibitem{LGGKM}
S.~Kharchev, A.~Marshakov, A.~Mironov and A.~Morozov,
 %``Landau-Ginzburg topological theories in the framework of GKM and equivalent
%hierarchies,''
Mod.\ Phys.\ Lett.\ A {\bf 8}, 1047 (1993)
%[Theor.\ Math.\ Phys.\  {\bf 95}, 571 (1993\ TMFZA,95,280-292.1993)]
[arXiv:hep-th/9208046].
%
\bibitem{VafOR}
A.~Okounkov, N.~Reshetikhin and C.~Vafa,
%``Quantum Calabi-Yau and classical crystals,''
arXiv:hep-th/0309208.
%
\bibitem{VafNekIqbal}
A.~Iqbal, N.~Nekrasov, A.~Okounkov and C.~Vafa,
%``Quantum foam and topological strings,''
arXiv:hep-th/0312022.
%
\bibitem{MZ}
J.~A.~Minahan and K.~Zarembo,
%``The Bethe-ansatz for N = 4 super Yang-Mills,''
JHEP {\bf 0303}, 013 (2003)
[arXiv:hep-th/0212208];\\
N.~Beisert, J.~A.~Minahan, M.~Staudacher and K.~Zarembo,
%``Stringing spins and spinning strings,''
JHEP {\bf 0309}, 010 (2003)
[arXiv:hep-th/0306139].


\end{thebibliography}
\end{document}